\documentclass[twocolumn,showpacs,preprintnumbers,amsmath,amssymb]{revtex4}

\usepackage{graphicx}
\usepackage{dcolumn}
\usepackage{bm}

\begin{document}

\preprint{APS/123-QED}

\title{Near-field interaction between domain walls in adjacent Permalloy nanowires}

\author{Liam O'Brien}
\author{D. Petit}%
\author{H. T. Zeng}
\author{E. R. Lewis}
\author{J. Sampaio}
\author{A. V. Jausovec}
\author{D.E Read}
\author{R. P. Cowburn}

\affiliation{%
Nanoscale Magnetics group, Department of Physics, Blackett Laboratory, Imperial College London, Prince Consort Road, London SW7 2AZ, United Kingdom
}%

\date{\today}

\begin{abstract}
The magnetostatic interaction between two oppositely charged transverse domain walls (DWs) in adjacent Permalloy nanowires is experimentally demonstrated. The dependence of the pinning strength on wire separation is investigated for distances between 13 and 125 nm, and depinning fields up to 93 Oe are measured. The results can be described fully by considering the interaction between the full complex distribution of magnetic charge within rigid, isolated DWs.
This suggests the DW internal structure is not appreciably disturbed by the pinning potential, and that they remain rigid although the pinning strength is significant.
This work demonstrates the possibility of non-contact DW trapping without DW perturbation and full continuous flexibility of the pinning potential type and strength. The consequence of the interaction on DW based data storage schemes is evaluated.  
\end{abstract}


\maketitle

Quantifying the interaction between magnetic domain walls (DWs) in closely packed networks of ferromagnetic nanowires is of vital importance to recently proposed DW based logic and data storage schemes \cite{allwd05,prkn08}, as these interactions could potentially compromise coherent DW propagation and correct device function.
In addition to these schemes, DWs in nanowires have also been suggested for use in a wide range of applications, such as atom trapping for quantum information processing \cite{allwood06}.
Furthermore, the fundamental properties of DWs themselves are attracting great interest, and DWs can be considered not only as transition regions between oppositely magnetized domains, but as quasi-particles \cite{saitoh04}. 
Their equilibrium \cite{nktn05}, dynamic properties \cite{beach05,thvll06,hysh07,yang08} and interactions with artificial defects \cite{petit08} are being intensively studied.
The ability to hold DWs at defined positions is required for a wide range of spin torque experiments, where current is used to depin the DWs; 
a continuously variable pinning strength which does not alter the current path is highly desirable in such cases \cite{hayashi06}.
The possibility of a well defined, tunable pinning potential where the DW acts as a truly rigid quasi-particle is also appealing, for example, in current induced resonance experiments \cite{bedau07, thomas06}. 
The effect of the presence of a transverse DW (TDW) in a nearby structure on the magnetic state of a ferromagnetic ring has been reported \cite{klaui05,lfnbrg06}, and quantified in terms of an additional local field due to the TDW stray field, but no quantitative analysis of the full pinning potential created by a DW on another DW has been made. 

In this paper, we experimentally study the interaction between two oppositely charged DWs travelling in two parallel Permalloy (Py) nanowires and its dependence on wire separation. 
The separations probed in this investigation are below the dimensions of the DW ($\sim 100nm$) itself and so are within the near-field limit. By considering the full, rigid magnetostatic charge distribution of an isolated DW we are able to reproduce fully the experimental results. This suggests the internal structure of the DW is not appreciably perturbed by the interaction for the geometries studied. 
This work opens up the possibility of continuously tailorable, remote DW pinning, which does not affect the DW internal structure.

\begin{figure}[h]
\centering
\includegraphics{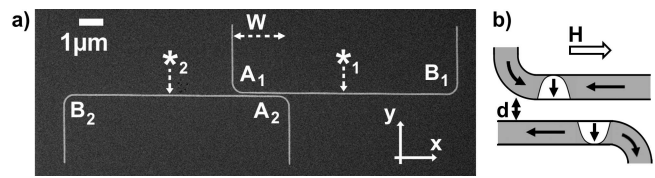}
\caption{
a) SEM image of a typical structure showing two $10\mu m$ long `U' shaped Py nanowires separated by an edge to edge distance $d=100$ nm. The asterisks indicate the positions where the MOKE laser spot is focused.
 b) Schematic of oppositely charged TDWs created in `$A$' corners moving, under field $H$, along wires separated by $d$.}
\label{figExptSetup}
\end{figure}

$100$ nm wide, $8$ nm thick, thermally deposited Py nanowires were fabricated using electron beam lithography (EBL) and a lift-off method. Fig.~\ref{figExptSetup}a shows a scanning electron microscope (SEM) image of a typical structure: two U-shaped nanowires with an edge to edge separation, $d$, between $125$ nm and $13$ nm, and vertical arm displacement, $w$, equal to $0.5$, $1$ or $1.5$ $\mu $m. Magnetic field sequences were applied in the plane of the sample at a frequency of 1 Hz using a quadrupole electromagnet. Two experiments were performed. The first was to create two oppositely charged DWs in the nearest pair of corners ($A_1$ and $A_2$). In this case (see Fig. \ref{figExptSetup}b), the DWs must pass within close proximity of one another before switching the remaining portion of each wire. The attractive interaction of the oppositely charged DWs will therefore directly influence the field required to separate them and switch the rest of the wire. In the second case, DWs are created in corners $B_1$ and $B_2$ and so are able to switch the central portion of each wire before meeting. This second setup acts as a control experiment to observe switching without any DW-DW interaction.

To create DWs at $A$ ($B$) corners a $45^{\circ}$ field pulse was applied in the $+x,+y$ ($-x,+y$) direction, giving an upwards tail to tail (TT) DW at corner $A_1$ ($B_1$) and upward head to head (HH) DW at corner $A_2$ ($B_2$). The field was then decreased to zero at 45$^{\circ}$, and a negative (positive) $x$ field was applied to move the two DWs towards one another. The sequence was repeated with all fields reversed to create a downward HH DW at $A_1$ ($B_1$) and a downward TT DW at $A_2$ ($B_2$), moving them towards one another with a positive (negative) $x$ field. 

\begin{figure}[h]
\centering
\includegraphics{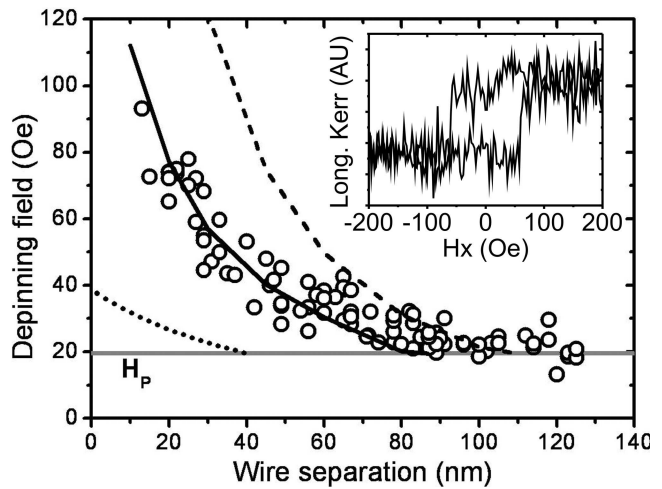}
\caption{Open circles: experimental dependence of DW transmission field $H_D$ on wire separation $d$. Grey line, average wire propagation field $H_P$. Modelled depinning fields with a fitting scale factor $\beta =0.54$, considering either a full rigid DW charge distribution (solid line), line-line interaction (dashed) or point-point interaction (dotted). Inset: Example MOKE observed hysteresis loop of a nanowire. Switching occurs at $H_D$, in this case $60$ Oe, due to depinning from the DW-DW potential well.}
\label{figExptResults}
\end{figure}

High sensitivity, spatially resolved magneto-optical Kerr effect (MOKE) measurements were used to probe the magnetization of the central portion of each wire (marked with $*$). Inset Fig. \ref{figExptResults} is an example of the observed hysteresis loop for the switching of a nanowire. The $60$ Oe switching field is due to the pinning from the DW-DW interaction. 
The switching fields of portions $*_{1,2}$ of the nanowires were measured as a function of $d$ (measured using SEM images). The results for 90 structures are shown in Fig.~\ref{figExptResults}. 
Switching of the nanowires in the control experiment (DWs created in corners $B$) occurs at an average propagation field $H_P$ of $19 \pm 4$ Oe (grey line). Fields $H_{D1}$ and $H_{D2}$ at which $*_1$ and $*_2$ reverse in the case where the DWs can interact (DWs created at corners $A$) are seen to increase as $d$ decreases. The mean difference between $H_{D1}$ and $H_{D2}$ is only $0.1$ Oe with a standard deviation of $2.1$ Oe; therefore only the average value $H_D$ for each structure has been plotted (open circles). We see a monotonic decrease of $H_{D}$ as $d$ increases from a maximum depinning field of $93$ Oe measured in wires with $d=13$ nm down to $\sim 19$Oe ($\sim H_P$) for $d\sim 90$ nm. Above $d\sim 90$ nm, the dependence changes as $H_D$ remains approximately constant at $H_P$. At these values of $d$ the dominant pinning mechanism is due to the inherent roughness of the nanowires and so the interaction of the DWs is no longer measured. No observable dependence was found on $w$. Separate studies, not presented here, ruled out a localised change in wire geometry causing the observed DW pinning \cite{prxmtyffct}, suggesting the observed $H_D$ is solely due to the interaction between the two DWs. This is also supported by the fact $H_{D1}$ and $H_{D2}$ are near identical for a pair of wires. The narrow distribution of $H_P$ (as measured from control experiment) shows we are not measuring a statistical variation of $H_P$ along these wires. 

To account for the DW-DW interaction dependence on $d$, we describe it in terms of the interaction between  magnetic charges, where the magnetic charge density $\rho _M$ is defined as $\rho_{M} = -\mu _{0} \bm{\nabla} \cdot \bm{M}$ and $\bm{M}$ is the local magnetization. We expect a model which considers the full DW charge distribution within the wire to match experimentally determined data if the two DWs remain rigid throughout the interaction and we account for finite temperature. 
OOMMF simulations \cite{OOMMF} ($3.5$ x $3.5$ x $4 \textrm{nm}^3$ cell size, $M_{s} = 800$ kA/m, $A=13\times 10^{-12}$ J/m and damping constant $\alpha= 0.5$) were used to calculate the magnetization configuration of the transverse DW (TDW) which is stable for our wire geometry ($100$ nm wide, $8$ nm thick) \cite{mcmichael97, laufenberg06}.
From this the charge distribution of the full DW \cite{chargedist} was obtained. The charge density is very inhomogeneously distributed, reflecting the triangular shape of TDWs \cite{mcmichael97}, with a positive surface charge along the wide edge of the DW, and a smaller negative charge along the narrow edge.

The interaction energy, $U$, between two extended charge distributions, $D_1$ and $D_2$, is given by:
\begin{equation}
U = \frac{1}{4\pi \mu _0} \sum_{i \in D_1}\sum_{j \in D_2}\frac{q_{i} q_{j}}{r_{ij}}  
\label{eqnU}
\end{equation} 
where $r_{ij}$ is the distance between the two charges $q_{i}$ and $q_{j}$.
In the case where the two distributions are associated with oppositely charged rigid DWs travelling in two adjacent wires separated by $d$ (see Fig. \ref{figExptSetup} b), $U$ is a potential well, and depends on their lateral separation $x$. 
The change in Zeeman energy, $\delta U_{Z}$ caused by separating the DWs by $\delta x$ under a field $H$ is $\delta U_{Z} = -2M_{s}HS \delta x$, where $M_{s}$ is the saturation magnetization and $S$ is the cross sectional area of the wire. 
The potential landscape $U(x)$ is sheared by a gradient equal to $-2M_{s}HS$, and the field required for DW depinning is therefore given by:
\begin{equation}
H_{D}= \frac{1}{2M_{s}S} \left. \frac{\partial U}{\partial x}\right| _{Max}
\label{eqnHDepin}
\end{equation}
The solid line in Fig. \ref{figExptResults} displays $H_D^\textrm{Full}$ (depinning fields for the full charge distribution) as a function of $d$, fitted to the experimental data (by least mean squares) using a  fitting scale factor $\beta =0.54$. Fitting to the data is excellent up to $d\simeq 90$ nm after which, as mentioned above, the inherent wire roughness becomes the dominant pinning mechanism.

This modelling does not include the effect of temperature; however, DW depinning is a thermally assisted process. At room temperature, depinning fields have previously been found to be between 40 and 65\% of the zero temperature simulated values \cite{jausovec07}. Ref. \cite{himeno05} experimentally shows similar reductions between measurements at 4.2 K and 300 K. The reduction in our modelling is well within the limits of these values. Our experimental results are therefore in very good agreement with a full rigid charge model,  
suggesting the interacting DWs can be considered as rigid quasi-particles with identical internal magnetic structure to that of a free, isolated DW. 

The far-field approximation of the charge distribution to a point-like volume charge is the simplest analytical model possible and may agree with the near-field experimental data depending on the homogeneity of the charge distribution. In order to test this, we consider a monopole containing the full DW charge $Q_\textrm{DW}= 2\mu_0M_sS$. A charge is located at the centre of each wire along the axis of symmetry of the DW, $w/2$ away from the wide edge where $w$ is the wire width. In this case, $H_{D}^P(d)$ is given by:
\begin{equation}
H_{D}^{P} = -\left(\frac{2}{3}\right)^{\frac{3}{2}}\frac{1}{K} \cdot \frac{Q_{DW}^2}{4\pi \mu_{0}}\cdot \frac{1}{(d+w)^2}
\end{equation}
The resulting $H_{D}^{P}$ vs $d$ curve is shown Fig. \ref{figExptResults} (lower dotted line), using $\beta=0.54$.  
The model is seen to underestimate $H_{D}$ over the region of $d$ investigated due to the artificial concentration of the charge at a single point, showing we cannot make a far-field approximation in this region of $d$ values and highlighting the inhomogeneity of the DW charge distribution. 
\begin{figure}[h]
\centering
\includegraphics{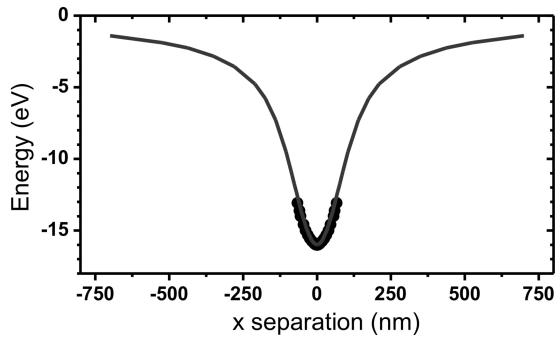}
\caption{Potential well obtained when laterally separating a pair of oppositely charged DWs by $x$ for the case $d = 30$ nm. Solid line: obtained considering the DW as a finite, rigid distribution of magnetostatic charges. Circles: obtained using OOMMF simulations.}
\label{figChargeResults}
\end{figure}

Without using a detailed knowledge of the charge distribution we could also attempt to form an analytical model based upon the geometry of the DW. TDWs have a triangular shape \cite{mcmichael97} as in Fig. \ref{figExptSetup}b with a large amount of the magnetization along the wide edge aligned in the $y$ direction. The charge due to the discontinuity of this magnetization across the wire edge will be significant and as the widest sides of the DWs are nearest we would expect this to be the overwhelming contribution to the interaction. In order to test this, we consider a uniform line of magnetic surface charge, magnitude $\sigma$ per unit length and length $2L$, with $\sigma \times 2L = Q_{DW}$ is used.   
The interaction energy $U^{L}$ between two such line charges of opposite sign, separated by $(x,d)$ is found by solving
\begin{equation}
U^{L}= -\frac{\sigma ^{2}}{4\pi \mu _{0}} \int_{-L}^L\left[ \int_{-L}^L \frac{1}{\sqrt{(x_1-x_2)^2+d^2}}\cdot dx_1 \right] \cdot dx_2 
\end{equation}    
$H_D^{L}$ for $U^L$ is shown Fig.\ref{figExptResults} (upper dashed line) for the case where $2L$ is taken as the DW width and $\beta = 0.54$. For this the DW width used is the FWHM of the $y$ component of magnetization along the wire centre in micromagnetic simulations ($85$ nm for our wire geometry).  
This attempt to use a more realistic analytical model without considering the detailed DW charge distribution is seen to overestimate the interaction strength.
A more complete analysis of the distribution in our wire geometry shows only $48\%$ of the DW charge is located along the wide edge, and has a Gaussian shape. 
Clearly then, positioning the full $Q_{DW}$ along the wire edge will overestimate the interaction strength. We conclude therefore that only by considering the near-field full charge distribution of the DW is it possible to closely match the experimentally observed magnetostatic interaction.

Micromagnetic simulations were performed (parameters as above), modelling the situation as in Fig. \ref{figExptSetup} b. 
The resulting depinning field dependence on $d$ is not shown, but is in good agreement with both experimental and near-field charge modelling (see Fig. \ref{figChargeResults}) and provides further evidence that little significant distortion occurs to the DW shape.   
\begin{figure}[h]
\centering
\includegraphics{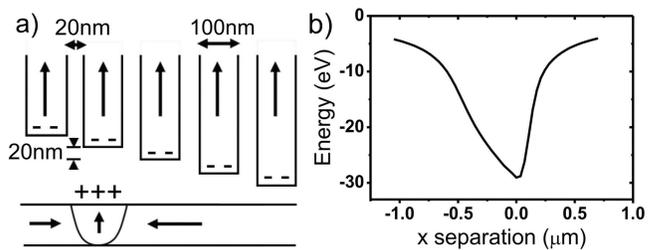}
\caption{Example of stray field based pinning creating an asymmetric potential well. a) Geometrical setup. b) Resulting asymmetric potential well experienced by the DW.
}
\label{figAsymmetricWell}
\end{figure}

Using one DW to pin another is a specific example of pinning by stray fields with little distortion to DW shape. It should therefore be possible to create a pinning potential of arbitrary shape with careful use of localised stray fields. Fig. \ref{figAsymmetricWell} shows an example of using multiple magnetic wire ends at different positions to create an asymmetric pinning potential. The depinning fields at $0$ K from each side of the well are $52$ and $168$ Oe respectively, which will act as a DW diode or ratchet similar to the geometrical protrusions in Refs. \cite{allwood04, himeno205}. By modifying the geometrical layout of such wires pinning strength and well shape can be varied. Using such a tailorable potential that does not alter wire or DW shape in spin torque and current induced resonant experiments will allow much clearer interpretation of current based effects.          

The interaction between DWs in neighboring nanowires is of great technological importance for DW based logic and data storage devices. Full charge modelling of DWs in nanowires of more technologically relevant dimensions ($45$ nm wide, 8nm thick, $45$ nm spacing), shows DW depinning at $71$ Oe (OOMMF simulations predict depinning at $67.5 \pm 2.5$ Oe) at $0$ K.  
This large interaction could result in cross talk errors between adjacent data channels.

In conclusion, we have experimentally studied the interaction between two DWs of opposite charge moving in parallel magnetic nanowires. The attractive interaction is found to result in depinning fields as high as $93$ Oe, measured in wires with separations of $13$ nm. 
Modelling using the full rigid magnetic charge distributions of isolated DWs reproduces the experimental results extremely well, suggesting that the DWs remain unperturbed by the interaction. Simple test analytical models that do not use the full charge distribution did not reproduce the experimental results, showing that consideration of the inhomogeneity of the charge is essential to understand near-field DW interactions. Pinning fields at $0$ K of $71$ Oe are expected in the technologically relevant dimensions of DW based data storage and logic devices.
This study experimentally demonstrates remote pinning using magnetostatic interaction with minimum perturbation to the DWs, and the possibility of continuous tailoring of pinning strength with no alteration to the nanowire.
This work was supported by the Corrigan Scholarship and the European Community under the Sixth Framework Programme SPINSWITCH (MRTN-CT-2006-035327).


\begin{thebibliography}{25}
\bibitem[{prx()}]{prxmtyffct}
\bibinfo{note}{EBL proximity effects can cause a local increase in wire width
  for structures with small $d$. Separate studies of pinning from similar
  defects give $H_{d}$ values significantly lower than those observed in this
  experiment.}
\end{thebibliography}
\end{document}